\documentclass[12pt]{article}
 \usepackage{graphicx}
 \usepackage[cp1251]{inputenc}

\begin{document}
 \noindent {\footnotesize\it
   Astronomy Letters, 2022, Vol. 48, No 1, pp. 9--19.}
 \newcommand{\dif}{\textrm{d}}

 \noindent
 \begin{tabular}{llllllllllllllllllllllllllllllllllllllllllllll}
 & & & & & & & & & & & & & & & & & & & & & & & & & & & & & & & & & & & & & &\\\hline\hline
 \end{tabular}

  \vskip 0.5cm
\centerline{\bf\large Kinematics of the Galaxy from Young Open Star Clusters}
\centerline{\bf\large with Data from the Gaia EDR3 Catalogue}
   \bigskip
   \bigskip
  \centerline
 {V. V. Bobylev \footnote [1]{e-mail: vbobylev@gaoran.ru} and A. T. Bajkova}
   \bigskip

  \centerline{\small\it Pulkovo Astronomical Observatory, Russian Academy of Sciences,}

  \centerline{\small\it Pulkovskoe sh. 65, St. Petersburg, 196140 Russia}
 \bigskip
 \bigskip
 \bigskip

{\bf Abstract}---We have analyzed the kinematics of open star clusters (OSCs) with the proper motions and distances calculated by Hao et al. based on Gaia EDR3 data. The mean line-of-sight velocities are known for a number of clusters from this list. We show that the Galactic rotation parameters determined from samples of OSCs with various ages are in good agreement between themselves. The most interesting results have been obtained from a sample of 967 youngest OSCs with a mean age of 18 Myr. In particular, we have found the following parameters of the angular velocity of Galactic rotation using only their proper motions and distances:
 $\Omega_0 =28.01\pm0.15$~km s$^{-1}$ kpc$^{-1},$
 $\Omega^{'}_0=-3.674\pm0.040$~km s$^{-1}$ kpc$^{-2},$ and
 $\Omega^{''}_0=0.565\pm0.023$~km s$^{-1}$ kpc$^{-3}$.
The circular rotation velocity of the solar neighborhood around the Galactic center here is $V_0=226.9\pm3.1$~km s$^{-1}$ for the adopted Galactocentric distance of the Sun $R_0=8.1\pm0.1$~kpc. The parameters of the spiral density wave have been determined from the space velocities of 233 young clusters. The amplitudes of the radial and tangential velocity perturbations produced by the spiral density wave are $f_R=9.1\pm0.8$~km s$^{-1}$ and $f_\theta=4.6\pm1.2$~km s$^{-1}$, respectively; the perturbation wavelengths are $\lambda_R=3.3\pm0.5$~kpc and $\lambda_\theta=2.6\pm0.6$~kpc for the the adopted four-armed spiral pattern. The Sun's phase in the spiral density wave has been found to be $(\chi_\odot)_R\approx-180^\circ$.


 \subsection*{INTRODUCTION}
Open star clusters (OSCs) are of great importance for studying the structure and kinematics of the Galaxy. In particular, they are used to estimate the parameters of the Galactic rotation curve (Glushkova et al. 1998; Zabolotskikh et al. 2002; Loktin and Beshenov 2003; Piskunov et al. 2006; Loktin and Popova 2019), the geometrical and kinematic characteristics
of the spiral density wave (Amaral and L\'epine 1997; Popova and Loktin 2005; Loktin and
Popova 2007; Naoz and Shaviv 2007; Bobylev et al. 2008; L\'epine et al. 2008; Junqueira et al. 2015; Camargo et al. 2015; Bobylev and Bajkova 2019; Cantat-Gaudin et al. 2020), and their other structural and kinematic properties (Babusiaux et al. 2018; Kuhn et al. 2018; Tarricq et al. 2021; Monteiro et al. 2021).

The number of discovered and studied OSCs increases steadily (Dias et al. 2001, 2006, 2021;
Kharchenko et al. 2005, 2007, 2013; Scholz et al. 2015; Cantat-Gaudin et al. 2018; Hao et al. 2021). The accuracy of their mean proper motions, line-of-sight
velocities, and distances improves.

The accuracy of the kinematic parameters of OSCs is of great importance for solving a great
variety of kinematic problems. The distances to OSCs are commonly estimated with the help of
the Hertzsprung–Russell diagram from photometric data or using other indirect methods (without trigonometric parallaxes). The implementation of the Gaia space experiment (Prusti et al. 2016) has made it possible not only to calculate highly accurate mean proper motions (Cantat-Gaudin et al. 2018) and line-of-sight velocities of OSCs, but also their mean trigonometric parallaxes (Cantat-Gaudin et al. 2020; Hao et al. 2021).

At present, the Gaia EDR3 (Gaia Early Data Release 3, Brown et al. 2021) version has been
published, where, in comparison with the previous Gaia DR2 version (Brown et al. 2018), the trigonometric parallaxes and proper motions were improved
approximately by 30\% for $\sim$1.5 billion stars. In the
Gaia EDR3 catalogue the trigonometric parallaxes
for $\sim$500 million stars were measured with errors less
than 0.2 milliarcseconds (mas), i.e., approximately
a third of the stars with measured parallaxes. The
proper motions for about a half of the stars in the
catalogue were measured with a relative error less
than 10\%.

A slight systematic offset with respect to an inertial reference frame apparently remains in the Gaia EDR3 parallaxes (Ren et al. 2021; Maiz Apell\'aniz 2021). This offset was first revealed in the Gaia DR2 parallaxes with $\Delta\pi=-0.029$~mas
(Lindegren et al. 2018). This correction should be added to the measured parallaxes and, therefore, the true distances to stars must slightly decrease.

For stars with magnitudes $G<15^m$ the random measurement errors of the proper motions lie within the range 0.02--0.04 mas yr$^{-1}$ (Brown et al. 2021), and they increase quite dramatically for fainter stars. There are no new line-of-sight velocity measurements
in the Gaia EDR3 catalogue. Thus, the line-of-sight velocities for more than 7 million stars are taken from the Gaia DR2 version.

The goal of this paper is to determine the Galactic rotation parameters and spiral density wave parameters based on the latest data on OSCs. For this
purpose, we use the mean proper motions and parallaxes
of OSCs calculated by Hao et al. (2021) based
on Gaia EDR3 data; the mean line-of-sight velocities
are also available for a number of clusters.

 \section*{METHOD}
We have three stellar velocity components from observations: the line-of-sight velocity $V_r$ and the two tangential velocity components $V_l=4.74r\mu_l\cos b$ and $V_b=4.74r\mu_b,$ along the Galactic longitude $l$ and latitude $b$ respectively, expressed in km s$^{-1}$. Here, 4.74 is the dimension coefficient and $r$ is the stellar heliocentric distance in kpc. The proper motion components $\mu_l\cos b$ and $\mu_b$ are expressed in mas yr$^{-1}$. The velocities $U,V,W$ directed along the rectangular Galactic coordinate axes are calculated via the components $V_r, V_l, V_b$:
 \begin{equation}
 \begin{array}{lll}
 U=V_r\cos l\cos b-V_l\sin l-V_b\cos l\sin b,\\
 V=V_r\sin l\cos b+V_l\cos l-V_b\sin l\sin b,\\
 W=V_r\sin b                +V_b\cos b,
 \label{UVW}
 \end{array}
 \end{equation}
where the velocity $U$ is directed from the Sun toward
the Galactic center, $V$ is in the direction of Galactic
rotation, and $W$ is directed to the north Galactic pole.
We can find two velocities, $V_R$ directed radially away
from the Galactic center and $V_{circ}$ orthogonal to it
pointing in the direction of Galactic rotation, based
on the following relations:
 \begin{equation}
 \begin{array}{lll}
  V_{circ}= U\sin \theta+(V_0+V)\cos \theta, \\
       V_R=-U\cos \theta+(V_0+V)\sin \theta,
 \label{VRVT}
 \end{array}
 \end{equation}
where the position angle $\theta$ obeys the relation $\tan\theta=y/(R_0-x)$, $x, y, z$ are the rectangular heliocentric coordinates of the star (the velocities $U, V, W$ are
directed along the corresponding $x, y, z$ axes), and $V_0$ is the linear rotation velocity of the Galaxy at the solar distance $R_0.$ The velocities $V_R$ and $W$ are virtually
independent of the pattern of the Galactic rotation curve. However, to analyze the periodicities in the tangential velocities, it is necessary to determine a
smoothed Galactic rotation curve and to form the residual velocities $\Delta V_{circ}$.

To determine the parameters of the Galactic rotation curve, we use the equations derived from Bottlinger's formulas, in which the angular velocity $\Omega$ is expanded into a series to terms of the second order of smallness in $r/R_0:$
\begin{equation}
 \begin{array}{lll}
 V_r=-U_\odot\cos b\cos l-V_\odot\cos b\sin l-W_\odot\sin b\\
 +R_0(R-R_0)\sin l\cos b\Omega^\prime_0
 +0.5R_0(R-R_0)^2\sin l\cos b\Omega^{\prime\prime}_0,
 \label{EQ-1}
 \end{array}
 \end{equation}
 \begin{equation}
 \begin{array}{lll}
 V_l= U_\odot\sin l-V_\odot\cos l-r\Omega_0\cos b\\
 +(R-R_0)(R_0\cos l-r\cos b)\Omega^\prime_0
 +0.5(R-R_0)^2(R_0\cos l-r\cos b)\Omega^{\prime\prime}_0,
 \label{EQ-2}
 \end{array}
 \end{equation}
 \begin{equation}
 \begin{array}{lll}
 V_b=U_\odot\cos l\sin b + V_\odot\sin l \sin b-W_\odot\cos b\\
 -R_0(R-R_0)\sin l\sin b\Omega^\prime_0
    -0.5R_0(R-R_0)^2\sin l\sin b\Omega^{\prime\prime}_0,
 \label{EQ-3}
 \end{array}
 \end{equation}
where $R$ is the distance from the star to the Galactic rotation axis, $R^2=r^2\cos^2 b-2R_0 r\cos b\cos l+R^2_0.$ The velocities $(U,V,W)_\odot$ are the mean group velocity
of the sample, are taken with the opposite sign, and reflect the peculiar motion of the Sun; $\Omega_0$ is the angular velocity of Galactic rotation at the solar distance
$R_0,$ the parameters $\Omega^{\prime}_0$ and $\Omega^{\prime\prime}_0$ are the corresponding derivatives of the angular velocity, and $V_0=R_0|\Omega_0|.$ The velocities $V_R$ and $\Delta V_{circ}$ must be freed from the peculiar solar velocity $U_\odot,V_\odot,W_\odot$. In this paper $R_0$ is taken to be $8.1\pm0.1$~kpc, according to the review by Bobylev and Bajkova (2021), where it was derived as a weighted mean of a large number of present-day individual estimates.

The influence of the spiral density wave in the radial ($V_R$) and residual tangential ($\Delta V_{circ}$) velocities is periodic with an amplitude $\sim6-10$ km s$^{-1}$. According to the linear theory of density waves (Lin and Shu 1964), it is described by the following relations:
 \begin{equation}
 \begin{array}{lll}
       V_R =-f_R \cos \chi,\\
 \Delta V_{circ}= f_\theta \sin\chi,
 \label{DelVRot}
 \end{array}
 \end{equation}
where
 \begin{equation}
 \chi=m[\cot(i)\ln(R/R_0)-\theta]+\chi_\odot
 \end{equation}
is the phase of the spiral density wave ($m$ is the number of spiral arms, $i$ is the pitch angle of the spiral pattern, and $\chi_\odot$ is the Sun's radial phase in the spiral
density wave); $f_R$ and $f_\theta$ are the amplitudes of the radial and tangential velocity perturbations, which are assumed to be positive. The periodicities associated
with the spiral density wave also manifest themselves
in the vertical velocities $W$ of young Galactic objects
(Bobylev and Bajkova 2015; Rastorguev et al. 2017).

We apply a modified spectral analysis (Bajkova and Bobylev 2012) to study the periodicities in the velocities $V_R$ and $\Delta V_{circ}$. The wavelength $\lambda$ (the
distance between adjacent spiral arm segments measured along the radial direction) is calculated from the relation
\begin{equation}
 2\pi R_0/\lambda=m\cot(|i|).
 \label{a-04}
\end{equation}
Let there be a series of measured velocities $V_{R_n}$ (these can be both radial ($V_R$) and tangential ($\Delta V_{circ}$) velocities), $n=1,\dots,N$, where $N$ is the number of
objects. The objective of our spectral analysis is to extract a periodicity from the data series in accordance with the adopted model describing a spiral density
wave with parameters $f,$ $\lambda$~(or $i)$ and $\chi_\odot$.

Having taken into account the logarithmic behavior of the spiral density wave and the position angles of objects $\theta_n$, our spectral (periodogram) analysis of the
series of velocity perturbations is reduced to calculating the square of the amplitude (power spectrum) of the standard Fourier transform (Bajkova and Bobylev 2012):
\begin{equation}
 \bar{V}_{\lambda_k} = \frac{1} {N}\sum_{n=1}^{N} V^{'}_n(R^{'}_n)
 \exp\biggl(-j\frac {2\pi R^{'}_n}{\lambda_k}\biggr),
 \label{29}
\end{equation}
where $\bar{V}_{\lambda_k}$ is the $k$th harmonic of the Fourier transform with wavelength $\lambda_k=D/k$, $D$ is the period of the series being analyzed,
 \begin{equation}
 \begin{array}{lll}
 R^{'}_{n}=R_0\ln(R_n/R_0),\\
 V^{'}_n(R^{'}_n)=V_n(R^{'}_n)\times\exp(jm\theta_n).
 \label{21}
 \end{array}
\end{equation}
The sought-for wavelength $\lambda$ corresponds to the peak
value of the power spectrum $S_{peak}.$ The pitch angle
of the spiral density wave is found from Eq. (8). We
determine the perturbation amplitude and phase by
fitting the harmonic with the wavelength found to the
observational data. The following relation can also be
used to estimate the perturbation amplitude:
 \begin{equation}
 f_R(f_\theta)=2\times\sqrt{S_{peak}}.
 \label{Speak}
 \end{equation}

\begin{figure}[t]
{ \begin{center}
  \includegraphics[width=0.95\textwidth]{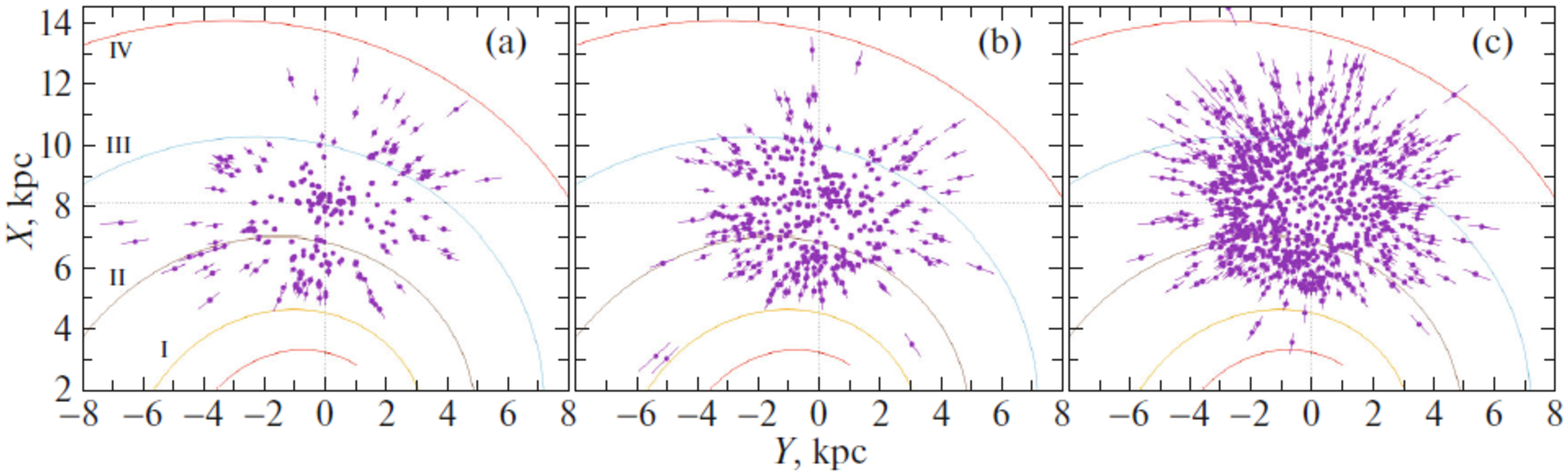}
  \caption{
Distribution of OSCs younger than 60 Myr (a), with ages in the interval 60--300 Myr (b), and older than 300 Myr (c) on the Galactic $XY$ plane; the four-armed spiral pattern with a pitch angle $i=-13^\circ$ (Bobylev and Bajkova 2014) is shown.}
 \label{f-XY-age}
\end{center}}
\end{figure}
 \begin{table}[t] \caption[]{\small
The Galactic rotation parameters found from OSCs of various ages only from their proper motions (Eqs. (4) and (5)), $N_\star$ is the number of clusters used, $\overline {t}$ is the mean age of the sample
 }
  \begin{center}  \label{t:01}
  \small
  \begin{tabular}{|l|r|r|r|r|r|}\hline
    Parameters              &      $<60$ Myr  &     $60-300$ Myr &    $>300$ Myr\\\hline
     $N_\star$              &            967  &             863  &          1794\\
   $\overline {t},$ Myr     &             18  &              163 &          1100\\
     ${\overline z},$ pc    &      $-20\pm4$  &        $-19\pm8$ &     $-24\pm9$\\
                                 &&&\\
    $U_\odot,$ km s$^{-1}$  &   $ 7.34\pm0.31$ & $ 7.64\pm0.44$ & $  8.64\pm0.35$\\
    $V_\odot,$ km s$^{-1}$  &   $10.61\pm0.45$ & $13.07\pm0.61$ & $ 17.67\pm0.46$\\
    $W_\odot,$ km s$^{-1}$  &   $ 7.45\pm0.23$ & $ 7.33\pm0.31$ & $  6.55\pm0.24$\\

  $\Omega_0,$ km s$^{-1}$ kpc$^{-1}$  & $28.01\pm0.15$ & $26.96\pm0.21$ & $ 27.06\pm0.16 $\\
  $\Omega^{'}_0,$ km s$^{-1}$ kpc$^{-2}$ & $-3.674\pm0.040$ & $-3.629\pm0.056$ & $-3.284\pm0.043$\\
 $\Omega^{''}_0,$ km s$^{-1}$ kpc$^{-3}$ & $ 0.565\pm0.023$ & $ 0.622\pm0.034$ & $ 0.463\pm0.021$\\

   $\sigma_0,$  km s$^{-1}$   &            7.0  &             9.1  &           10.3\\

   $A,$ km s$^{-1}$ kpc$^{-1}$ & $ 14.88\pm0.16$ &  $ 14.70\pm0.23$ &  $ 13.30\pm0.17$\\
   $B,$ km s$^{-1}$ kpc$^{-1}$ & $-13.13\pm0.22$ &  $-12.26\pm0.31$ &  $-13.76\pm0.24$\\
   $V_0,$  km s$^{-1}$   & $ 226.9\pm3.1$  &  $ 218.4\pm3.2$  &  $ 219.2\pm3.0$ \\
  \hline
 \end{tabular}\end{center} \end{table}
 \begin{table}[t] \caption[]{\small
The Galactic rotation parameters found from OSCs of various ages through the simultaneous LSM solution of the system of three equations (3)--(5), $N_\star$ is the number of clusters used, and NRV is the number of OSCs with the line-of-sight velocities
 }
  \begin{center}  \label{t:02}
  \small
  \begin{tabular}{|l|r|r|r|r|r|}\hline
    Parameters                 &  $<60$ Myr  & $60-300$ Myr & $>300$ Myr\\\hline
     $N_\star$                   &            967  &             863  &           1794\\
     $N_{RV}$                    &            233  &             398  &           1000\\
                                 &&&\\
    $U_\odot,$ km s$^{-1}$   &   $ 7.52\pm0.33$ & $ 8.57\pm0.43$ & $  9.70\pm0.38$\\
    $V_\odot,$ km s$^{-1}$   &   $12.43\pm0.45$ & $13.58\pm0.56$ & $ 19.80\pm0.46$\\
    $W_\odot,$ km s$^{-1}$   &   $ 7.53\pm0.27$ & $ 7.37\pm0.37$ & $  6.78\pm0.33$\\

  $\Omega_0,$ km s$^{-1}$ kpc$^{-1}$ & $28.07\pm0.16$ & $27.34\pm0.22$ & $ 27.62\pm0.19 $\\
  $\Omega^{'}_0,$ km s$^{-1}$ kpc$^{-2}$ & $-3.713\pm0.041$ & $-3.782\pm0.053$ & $-3.486\pm0.045$\\
 $\Omega^{''}_0,$ km s$^{-1}$ kpc$^{-3}$ & $ 0.613\pm0.025$ & $ 0.674\pm0.036$ & $ 0.548\pm0.024$\\

   $\sigma_0,$ km s$^{-1}$    &            8.2  &            10.9  &           13.9\\

   $A,$ km s$^{-1}$ kpc$^{-1}$ & $ 15.04\pm0.17$ &  $ 15.32\pm0.21$ &  $ 14.12\pm0.18$\\
   $B,$ km s$^{-1}$ kpc$^{-1}$ & $-13.03\pm0.23$ &  $-12.02\pm0.31$ &  $-13.50\pm0.26$\\
   $V_0,$ km s$^{-1}$       & $ 227.3\pm3.1$  &  $ 221.5\pm3.3$  &  $ 223.7\pm3.2$ \\
  \hline
 \end{tabular}\end{center} \end{table}

 \section*{DATA}
Here the paper by Hao et al. (2021), where the
mean proper motions and mean parallaxes of OSCs
were calculated based on Gaia EDR3 data, served
as the main data source. The age estimates were
collected by these authors from various sources. The
catalogue contains data on 3794 OSCs and, therefore,
it is the most extensive kinematic database of
Galactic OSCs to date.

The catalogue by Hao et al. (2021) provides the parallaxes $\pi$ via which below we calculate the distances $r$ from the formula $r=1/\pi.$ The relative errors
of the mean OSC parallaxes in the entire catalogue
are small, being, on average, $\sim$10\%. There are isolated
cases with parallax errors greater than 30\%, and
such OSCs are not used in this paper.

Figure 1 shows the distribution of OSCs from three samples of various ages in projection onto the Galactic $XY$ plane. We use the coordinate system in which the $X$ axis is directed from the Galactic center to the Sun and the direction of the $Y$ axis
coincides with the direction of Galactic rotation. The
four-armed spiral pattern with a pitch angle $i=-13^\circ$
(Bobylev and Bajkova 2014) constructed with $R_0=8.1$~kpc is shown; the Roman numerals number the following spiral arm segments: Scutum (I) , Carina-Sagittarius (II), Perseus (III), and the Outer Arm (IV).

The sample of OSCs younger than 60 Myr contains
a total of 967 members with a mean age
of 18 Myr. Figure 1a shows the distribution of
233 OSCs for which the line-of-sight velocities are
available. Based on the space velocities of these clusters, we perform a spectral analysis to determine the parameters of the spiral density wave.

The sample of OSCs with ages in the interval 60--300 Myr contains a total of 863 members. Here, the mean age of the clusters is 163 Myr. Figure 1b shows the distribution of 398 OSCs for which the line-of-sight  velocities are available.

The sample of OSCs older than 300 Myr contains
a total of 1794 members. The mean age of the
clusters in this sample is 1.1 Gyr. Figure~1c shows
1000 OSCs with the line-of-sight velocities.

Note that the distributions of all OSCs in projection onto the $XY$ plane divided into four age intervals are shown in Fig.~1 from Hao et al. (2021). It can be clearly seen there that the OSCs younger than 20~Myr and those in the interval 20--200~Myr have
a strong concentration toward the spiral arms.

\begin{figure}[t]
{ \begin{center}
  \includegraphics[width=0.7\textwidth]{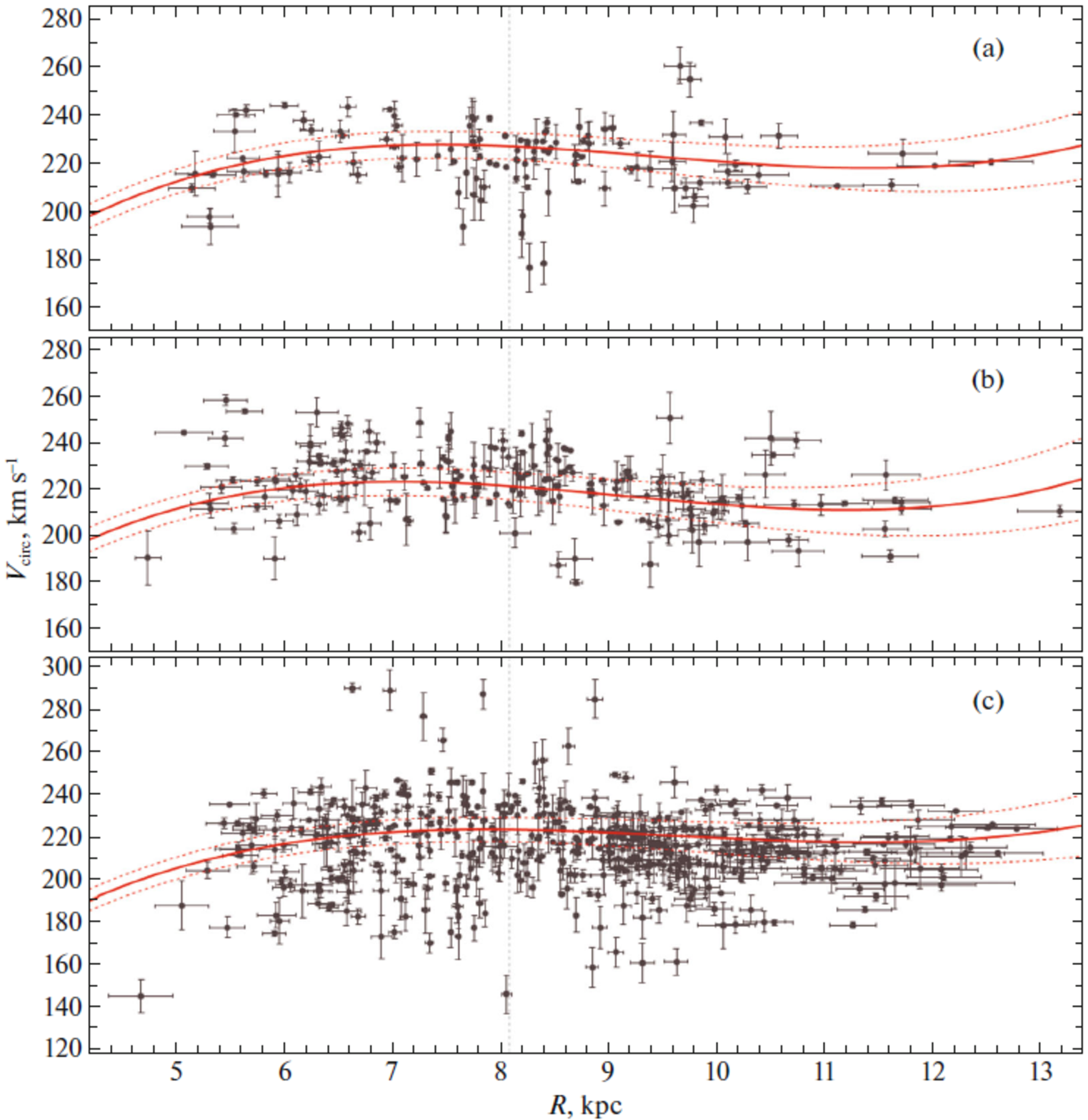}
  \caption{
Circular rotation velocities $V_{circ}$ versus distance $R$ for the youngest OSCs (a), with ages in the interval 60--300 Myr (b), and older than 300 Myr (c); the Galactic rotation curve found from these OSCs with an indication of the boundaries of the 1$\sigma$
confidence regions is shown for each sample.}
 \label{f-2}
\end{center}}
\end{figure}
\begin{figure}[t]
{ \begin{center}
  \includegraphics[width=0.85\textwidth]{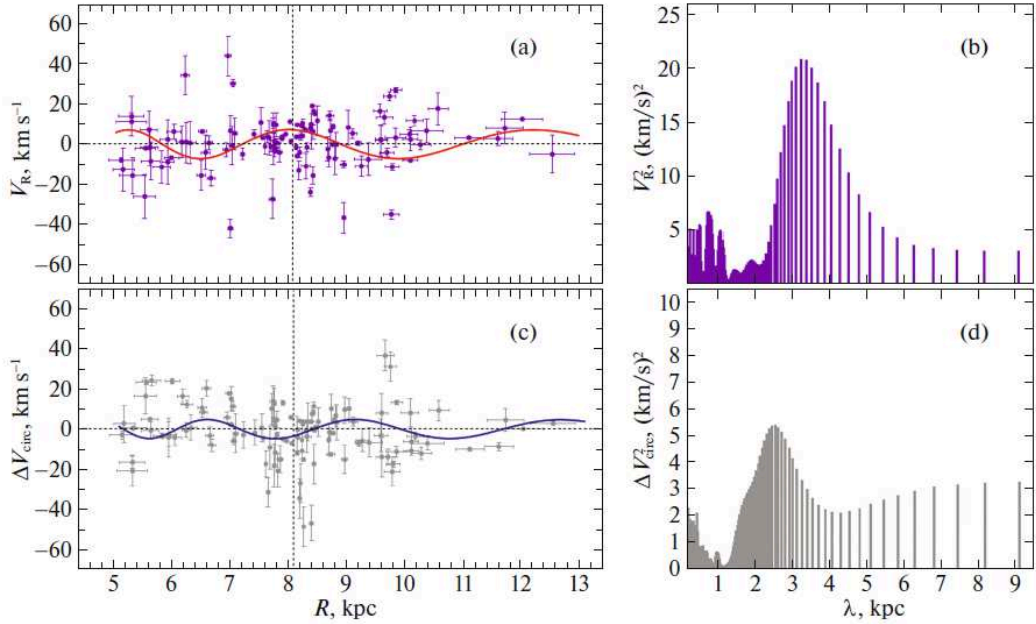}
  \caption{
Radial velocities $V_R$ versus distance $R$ for the youngest OSCs (a), the power spectrum of this sample (b), the residual rotation velocities $\Delta V_{circ}$ of the youngest OSCs (c), and their power spectrum (d).}
 \label{f-3}
\end{center}}
\end{figure}
\begin{figure}[t]
{ \begin{center}
  \includegraphics[width=0.85\textwidth]{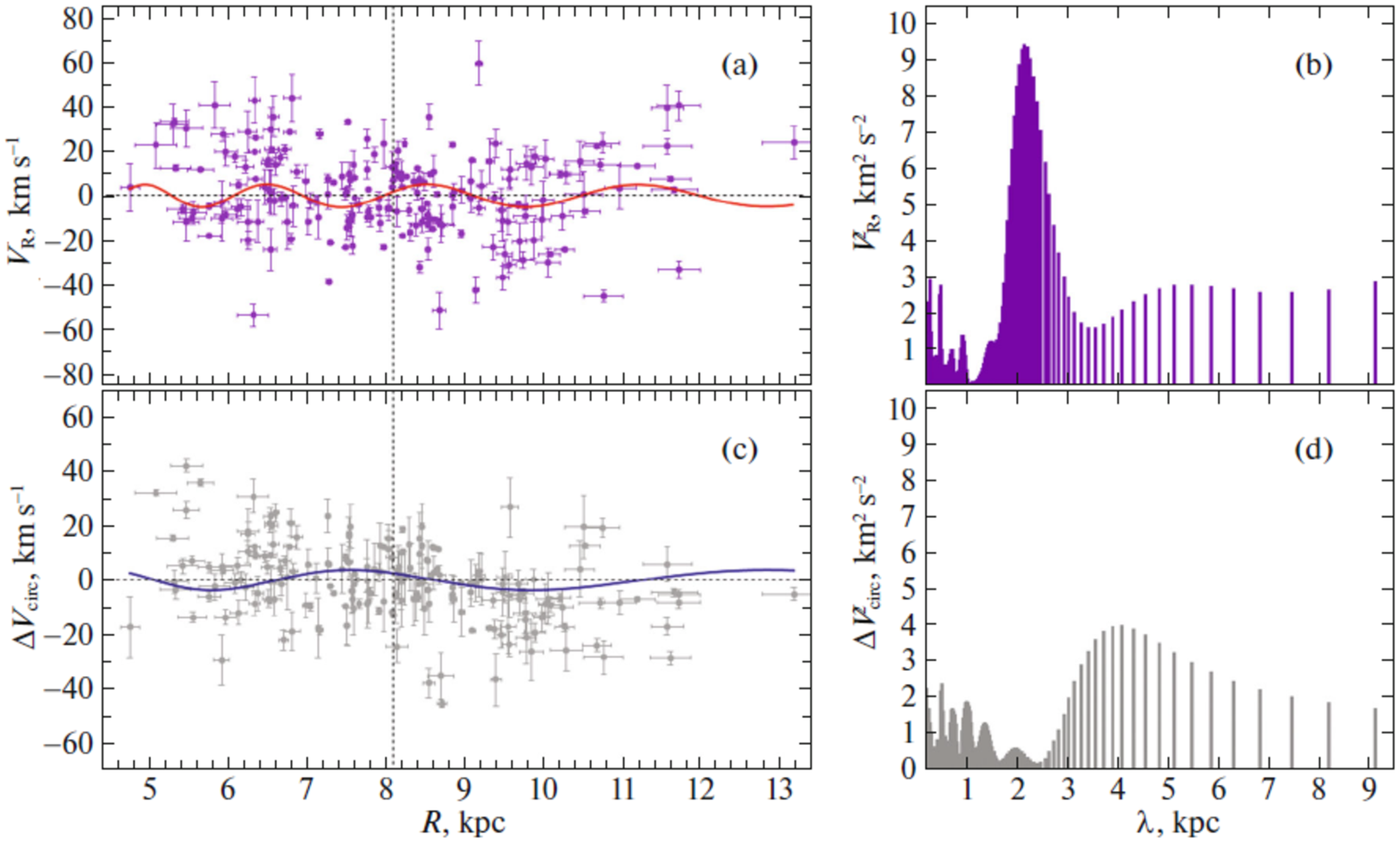}
  \caption{
Radial velocities $V_R$ versus distance $R$ for OSCs with ages in the interval 60--300 Myr (a), the power spectrum of this sample (b), the residual rotation velocities $\Delta V_{circ}$ for OSCs with ages in the interval 60--300 Myr (c), and their power spectrum (d).}
 \label{f-4}
\end{center}}
\end{figure}

 \section*{RESULTS}
The system of conditional equations (3)--(5) is solved by the least-squares method (LSM) with weights of the form $w_r=S_0/\sqrt {S_0^2+\sigma^2_{V_r}},$
 $w_l=S_0/\sqrt {S_0^2+\sigma^2_{V_l}}$ and
 $w_b=S_0/\sqrt {S_0^2+\sigma^2_{V_b}},$ where $S_0$
is the ``cosmic'' dispersion, $\sigma_{V_r}, \sigma_{V_l}, \sigma_{V_b}$ are the
dispersions of the corresponding observed velocities.
$S_0$ is comparable to the root-mean-square residual $\sigma_0$
(the error per unit weight) in solving the conditional
equations (3)--(5). We adopted $S_0=7-8$~km s$^{-1}$
when analyzing the sample of young OSCs and
$S_0=11-14$~km s$^{-1}$ for the sample of older OSCs.
The system of equations (3)--(5) was solved in several
iterations using the $3\sigma$ criterion to eliminate the
OSCs with large residuals.

 \subsection*{Method I}
The first method consists in seeking a solution based only on the OSC proper motions. In this case, the system of two conditional equations (4) and (5) is
solved.

The Galactic rotation parameters found for three samples of various ages are given in Table~1. For each sample we calculated the mean age $t$ and the mean coordinate $z$ (reflects the Sun's elevation above the Galactic plane). Note that the values of $z$ found are in excellent agreement with $z=-23\pm3$~pc found by analyzing OSCs with data from the Gaia DR2 catalogue in Cantat-Gaudin et al. (2020).

The Oort constants $A=0.5\Omega'_0 R_0$ and $B=A-\Omega_0$ calculated using the deduced
$\Omega_0$ and $\Omega'_0$ are given in the lower part of the table. The linear Galactic
rotation velocity at the solar distance $V_0=R_0|\Omega_0|$ for
the adopted $R_0=8.1\pm0.1$~kpc is also given.

Based on the entire sample of 3624 OSCs, we found the velocity components $(U,V,W)_\odot=(7.89,14.35,6.97)\pm(0.22,0.29,0.16)$~km s$^{-1}$ and
the following parameters of the angular velocity of
Galactic rotation by this method:
 \begin{equation}
 \label{solution-I}
 \begin{array}{lll}
      \Omega_0=~27.37\pm0.10~\hbox{km s$^{-1}$ kpc$^{-1}$},\\
  \Omega^{'}_0=-3.510\pm0.027~\hbox{km s$^{-1}$ kpc$^{-2}$},\\
 \Omega^{''}_0=~0.534\pm0.014~\hbox{km s$^{-1}$ kpc$^{-3}$}.
 \end{array}
 \end{equation}
In this solution the error per unit weight is $\sigma_0=9.4$~km s$^{-1}$. The linear Galactic rotation velocity at the solar distance is $V_0=221.7\pm2.9$~km s$^{-1}$, while
the Oort constants are $A=14.21\pm0.11$~km s$^{-1}$ kpc$^{-1}$
and $B=-13.15\pm0.15$~km s$^{-1}$ kpc$^{-1}$.

 \subsection*{Method II}
In this approach we use all of the available data.
The clusters with the proper motions, line-of-sight
velocities, and distances give all three equations (3)--(5), while the clusters for which only the proper motions are available give only two equations, (4) and
(5). We solve this system of equations simultaneously.

The Galactic rotation parameters found by this
method for three samples of OSCs with various ages
are presented in Table 2. The number of OSCs with
the line-of-sight velocities $N_{RV}$ used in the solution
is given. At the same time, the OSCs with errors
in their mean line-of-sight velocities greater than
10~km s$^{-1}$ were not used.

As a result of using the data on all 3624 OSCs, we found $(U,V,W)_\odot=(8.73,16.03,7.10)\pm(0.24,0.30,0.20)$~km s$^{-1}$ and
 \begin{equation}
 \label{solution-II}
 \begin{array}{lll}
      \Omega_0 =~27.79\pm0.12~\hbox{km s$^{-1}$ kpc$^{-1}$},\\
  \Omega^{'}_0 =-3.669\pm0.028~\hbox{km s$^{-1}$ kpc$^{-2}$},\\
 \Omega^{''}_0 =~0.606\pm0.016~\hbox{km s$^{-1}$ kpc$^{-3}$}.
 \end{array}
 \end{equation}
In this solution the error per unit weight is $\sigma_0=12.0$~km s$^{-1}$. The linear Galactic rotation velocity at the solar distance is $V_0=225.1\pm2.9$~km s$^{-1}$, while
the Oort constants are $A=14.86\pm0.11$~km s$^{-1}$ kpc$^{-1}$
and $B=-12.93\pm0.16$~km s$^{-1}$ kpc$^{-1}$.

As can be seen from a comparison of the parameters
(12) and (13) as well as Tables 1 and 2, invoking
the line-of-sight velocities leads to an increase in the
dispersion of the estimates.

In Fig. 2 the circular rotation velocities $V_{circ}$ are
plotted against the distance $R$ for three samples
of OSCs with various ages. The parameters from
the corresponding column in Table 2 were taken to
construct the rotation curve for each sample. We
see good agreement between these Galactic rotation
curves. Therefore, any of them can be used to form
the residual velocities $\Delta V_{circ}$ for a further spectral
analysis.

Note that we deem the Galactic rotation curve
obtained with the smallest error per unit weight $\sigma_0=7$~km s$^{-1}$ to be the best one. The parameters of this rotation curve found by method~I only from the proper
motions of the youngest OSCs are given in the first column of Table~1.

 \subsection*{Spectral Analysis}
First we determined the parameters of the spiral
density wave based on the sample of youngest OSCs
with ages less than 60 Myr (with a mean age of
18 Myr). For this purpose, we used 233 OSCs for
which the line-of-sight velocities are available. A
spectral analysis of their radial and residual tangential
velocities showed that the perturbation wavelengths
and velocity perturbations found independently for
each type of velocities agree in principle.

The results of our spectral analysis of the OSCs
from this sample are presented in Fig. 3. The figure
shows the radial velocities $V_R$ and residual rotation
velocities $\Delta V_{circ}$ as a function of distance R and their
power spectra.

Based on 233 OSCs from this sample, we found the wavelengths $\lambda_R=3.3\pm0.5$~kpc and $\lambda_\theta=2.6\pm0.6$~kpc. For the four-armed spiral pattern ($m=4$ and
the adopted $R_0$) the pitch angles $i_R=-14.5\pm2.1^\circ$ and $i_\theta=-11.4\pm2.6^\circ$ correspond to these values. The Sun’s phase in the spiral density wave is $(\chi_\odot)_R\approx-180^\circ$  and $(\chi_\odot)_\theta\approx-120^\circ$; we measure it from the presumed center of the Carina–Sagittarius arm--from $R\sim7$~kpc in the direction of increasing $R.$ The amplitudes of the radial and tangential velocity perturbations are $f_R=9.1\pm0.8$~km s$^{-1}$ and $f_\theta=4.6\pm1.2$~km s$^{-1}$, respectively.

A spectral analysis of the space velocities of OSCs with ages in the interval 60--300 Myr showed that there is also an influence of the spiral density wave in them. The results of our spectral analysis of the OSCs from this sample are presented in Fig. 4, where
the velocities $V_R$ and $\Delta V_{circ}$ and their power spectra are shown.

Based on 398 OSCs from this sample, we found the perturbation wavelengths $\lambda_R=2.2\pm0.6$~kpc and $\lambda_\theta=4.1\pm0.8$~kpc. For the four-armed spiral
pattern ($m=4$ and the adopted $R_0$) the pitch angles
$i_R=-9.6\pm2.6^\circ$ and $i_\theta=-17.9\pm3.3^\circ$ correspond
to these values. The amplitudes of the radial and
tangential velocity perturbations are $f_R=6.1\pm1.8$~km s$^{-1}$
and $f_\theta=3.9\pm2.2$~km s$^{-1}$, respectively.
The Sun’s phase in the spiral density wave $\chi_\odot$ here is
$(\chi_\odot)_R\approx-90^\circ$ and $(\chi_\odot)_\theta\approx-180^\circ$. We see that the parameters of the spiral density wave are determined from the radial velocities
of these OSCs quite reliably and in agreement with
the results described above.

 \section*{DISCUSSION}
 \subsection*{Velocities $(U,V,W)_\odot$}
The velocities $(U,V,W)_\odot$ are the group velocity of the OSC sample under consideration taken with the opposite sign. These velocities contain the peculiar motion of the Sun relative to the local standard of rest, the perturbations from the spiral density wave (for
relatively young objects), and the influence of the so-called asymmetric drift (lagging behind the circular rotation velocity with sample age) on the velocity $V_\odot$.

The components of the peculiar solar velocity relative to the local standard of rest are currently believed to have been determined most reliably in Sch\"onrich et al. (2010),
 $(U,V,W)_\odot=(11.1,12.2,7.3)\pm(0.7,0.5,0.4)$~km s$^{-1}$. We can see that the velocities $U_\odot$ and $W_\odot$ found in this paper from various OSC samples agree, within the error limits, with the estimates by Sch\"onrich et al. (2010). In
addition, an increase in the velocity $V_\odot$ with OSC age
can be seen in our results, which is a manifestation of
the asymmetric drift.

 \subsection*{Galactic Rotation}
The most important local parameter is the linear velocity $V_0$. Such objects of the Galactic thin disk as hydrogen clouds, maser sources, O- and B-type stars, young OSCs, Cepheids, etc. possess the most rapid rotation.

For example, Bobylev et al. (2016) obtained an estimate of $V_0=236\pm6$~km s$^{-1}$ for the adopted $R_0=8.3\pm0.2$~kpc by analyzing a sample of OSCs younger than 50~Myr from the Milky Way Star Clusters (MWSC) catalogue (Kharchenko et al. 2013).
While analyzing about 770 Cepheids with known
line-of-sight velocities, Mr\'oz et al. (2019) obtained
an estimate of $V_0=233.6\pm2.8$~km s$^{-1}$ for the
adopted $R_0=8.122\pm0.031$~kpc. Using about 3500
classical Cepheids, Ablimit et al. (2020) constructed
the Galactic rotation curve in the range of distances
$R:$ 4--19 kpc and found the velocity $V_0=232.5\pm0.9$~km s$^{-1}$ for the adopted $R_0=8.122\pm0.031$~kpc with a very high accuracy. Having analyzed 800 Cepheids with known line-of-sight velocities, Bobylev et al. (2021) found $V_0=240\pm3$~km s$^{-1}$ for
the inferred $R_0=8.27\pm0.10$~kpc.

Note several determinations of the parameters of the angular velocity of Galactic rotation using various data. For example, based on 130 Galactic
masers with measured trigonometric parallaxes, Rastorguev
et al. (2017) obtained the following estimates:
$(U,V)_\odot=(11.40,17.23)\pm(1.33,1.09)$~km s$^{-1}$,
 $\Omega_0=28.93\pm0.53$~km s$^{-1}$ kpc$^{-1}$,
 $\Omega^{'}_0=-3.96\pm0.07$~km s$^{-1}$ kpc$^{-2}$ and
$\Omega^{''}_0=0.87\pm0.03$~km s$^{-1}$ kpc$^{-3}$,
where the linear velocity $V_0$ is $243\pm10$~km s$^{-1}$ for
$R_0=8.40\pm0.12$~kpc found.

Based on a sample of 495 OB stars with their proper motions from the Gaia DR2 catalogue (Brown et al. 2018), Bobylev and Bajkova (2018) found the
following parameters: $(U,V,W)_\odot=(8.16,11.19,8.55)\pm(0.48,0.56,0.48)$~km s$^{-1}$,
      $\Omega_0=28.92\pm0.39$~km s$^{-1}$ kpc$^{-1}$,
  $\Omega^{'}_0=-4.087\pm0.083$~km s$^{-1}$ kpc$^{-2}$ and
 $\Omega^{''}_0=0.703\pm0.067$~km s$^{-1}$ kpc$^{-3}$,
 where $V_0=231\pm5$~km s$^{-1}$ for adopted $R_0=8.0\pm0.15$~kpc.

Based on a sample of 788 Cepheids from the list
of Mr\'oz et al. (2019) with their proper motions and
line-of-sight velocities from the Gaia DR2 catalogue,
Bobylev et al. (2021) found $(U_\odot,V_\odot,W_\odot)=(10.1,13.6,7.0)\pm(0.5,0.6,0.4)$~km s$^{-1}$, and:
      $\Omega_0=29.05\pm0.15$~km s$^{-1}$ kpc$^{-1}$,
   $\Omega^{'}_0=-3.789\pm0.045$~km s$^{-1}$ kpc$^{-2}$,
  $\Omega^{''}_0=0.722\pm0.027$~km s$^{-1}$ kpc$^{-3}$,
 $\Omega^{'''}_0=-0.087\pm0.007$~km s$^{-1}$ kpc$^{-4}$, $R_0=8.27\pm0.10$~kpc.

Based on a sample of 147 masers, Reid et al. (2019) found the following values of the two most important kinematic parameters: $R_0=8.15\pm0.15$~kpc and
$\Omega_\odot=30.32\pm0.27$~km s$^{-1}$ kpc$^{-1}$, where
 $\Omega_\odot=\Omega_0+V_\odot/R.$ The velocity $V_\odot=12.24$~km s$^{-1}$ was
taken from Sch\"onrich et al. (2010). These authors used an expansion of the linear Galactic rotation velocity into a series.

The Oort constants $A$ and $B$ are also of interest. For example, having analyzed the proper motions and parallaxes of a local sample of 304\,267
main-sequence stars from the Gaia DR1 catalogue
(Brown et al. 2016), Bovy (2017) found
$A=15.3\pm0.5$~km s$^{-1}$ kpc$^{-1}$ and $B=-11.9\pm0.4$~km s$^{-1}$ kpc$^{-1}$,
based on which he estimated the angular velocity
of Galactic rotation $\Omega_0=27.1\pm0.5$~km s$^{-1}$ kpc$^{-1}$
and the velocity $V_0=219\pm4$~~km s$^{-1}$. Based on a
sample of 5627 A-type stars close ($r<0.6$~kpc) to
the Sun from the LAMOST DR4 (The Large Sky
Area Multi-Object Fiber Spectroscopic Telescope)
catalogue (Cui et al. 2012; Xiang et al. 2017), Wang
et al. (2021) obtained the following estimates of the
Oort constants: $A=16.31\pm0.89$~km s$^{-1}$ kpc$^{-1}$
and $B=-11.99\pm0.79$~km s$^{-1}$ kpc$^{-1}$, where
$\Omega_0=28.30\pm1.19$~km s$^{-1}$ kpc$^{-1}$.

It can be concluded that the velocity $V_0$ found in
this paper from the youngest OSCs is in excellent
agreement with the estimates of this velocity obtained
from other young objects of the Galactic disk. The
parameters $\Omega_0$, $\Omega^{'}_0$ and $\Omega^{'}_0$, as well as the Oort
constants $A$ and $B$ were determined in this paper with a high accuracy; their values are also in good agreement with the estimates of other authors.

 \subsection*{Spiral Density Wave Parameters}
Mel’nik et al. (2001) found $f_R=7\pm1$~km s$^{-1}$, $f_\theta=2\pm1$~km s$^{-1}$, and $\lambda=2.0\pm0.2$~kpc for $m=2$ by analyzing OB associations. Zabolotskikh
et al. (2002) found $f_R=7\pm2$~km s$^{-1}$, $f_\theta=1\pm2$~km s$^{-1}$, and $i=-6.0\pm0.9^\circ$ for $m=2$ with a phase $\chi_\odot\approx-85^\circ$ based on young Cepheids ($P>9^d$) and OSCs ($\log t<7.6$); $f_R=6.6\pm2.5$~km s$^{-1}$, $f_\theta=0.4\pm2$~km s$^{-1}$, and $i=-6.6\pm0.9^\circ$ for $m=2$ with a
phase $\chi_\odot\approx-97^\circ$ based on OB stars.

Having analyzed the spatial distribution of a large sample of classical Cepheids, Dambis et al. (2015) estimated the pitch angle of the spiral pattern $i=-9.5\pm0.1^\circ$ and the Sun’s phase $\chi_\odot=-121\pm3^\circ$ for the four-armed spiral pattern.

Based on a sample of OSCs younger than 50 Myr from the MWSC catalogue (Kharchenko et al. 2013), Bobylev et al. (2016) found $f_\theta=5.6\pm1.6$~km s$^{-1}$ and $f_R=7.7\pm1.4$~km s$^{-1}$, the perturbation wavelengths $\lambda_\theta=2.6\pm0.5$~kpc ($i_\theta=-11\pm2^\circ$) and $\lambda_R=2.1\pm0.5$~kpc ($i_R=-9\pm2^\circ$) for the adopted four-armed spiral pattern ($m=4$).

Having analyzed maser sources with VLBI parallaxes, Rastorguev et al. (2017) found $i=-10.4\pm0.3^\circ$ and $\chi_\odot=-125\pm10^\circ,$ in good agreement with our results.

Loktin and Popova (2019) found $f_R=4.6\pm0.7$~km s$^{-1}$ and $f_\theta=1.1\pm0.4$~km s$^{-1}$ based on OSCs from the Homogenous Catalogue of Open Cluster Parameters with their proper motions from the Gaia DR2 catalogue. A review of the determinations
of the velocity perturbations $f_R$ and $f_\theta$ made in
recent years by various authors using various spiral
structure indicators can be found in the paper of these authors.

While analyzing 326 young ($\log t<8$) OSCs with the proper motions and distances calculated from Gaia DR2 data, Bobylev and Bajkova (2019) obtained
the following estimates: $f_\theta=3.8\pm1.2$~km s$^{-1}$ and
$f_R = 4.7\pm1.0$~km s$^{-1}$, $\lambda_\theta=2.3\pm0.5$~kpc and $\lambda_R=2.2\pm0.5$~kpc ($m=4, R_0=8.0\pm0.15$~kpc), and $\chi_\odot=-120^\circ\pm10^\circ$.

It can be noted that the amplitude of the tangential
perturbations $f_\theta$ is usually determined poorly. As
simulations of density waves in the Galaxy showed
(Burton 1971), the expected perturbation amplitudes
at the solar distance can reach $f_R\sim8$~km s$^{-1}$ and
$f_\theta\sim6$~km s$^{-1}$. We see that $f_R=9.1\pm0.8$~km s$^{-1}$
found in this paper from the youngest OSCs is in
excellent agreement with the expected estimate.

Based on OSCs with a mean age of 163 Myr, the parameters of the spiral density wave are well determined from the radial velocities. The Sun’s phases $(\chi_\odot)_R$ are of greatest interest here: $-180^\circ$ and $-90^\circ$ for OSCs with a mean age of 18 and 163 Myr, respectively, which show that the wave moves.

 \section*{CONCLUSIONS}
We studied a sample of OSCs with their proper motions and parallaxes from the Gaia EDR3 catalogue. The catalogue by Hao et al. (2021), which contains data on 3794 OSCs with various ages, served for this purpose. The mean line-of-sight velocities are
known approximately for a third of the clusters from this catalogue.

We showed that the Galactic rotation parameters determined from samples of OSCs with various
ages are in good agreement between themselves; the methods of analysis using the space velocities and only the proper motions of OSCs were applied. In particular, the linear rotation velocity of the solar neighborhood $V_0$ varies from $218\pm3$~km s$^{-1}$ found
from relatively old OSCs to $227\pm3$~km s$^{-1}$ typical for the youngest OSCs.

We analyzed in detail the kinematics of 967 youngest OSCs with a mean age of 18 Myr. Primarily these OSCs were used to redetermine the Galactic rotation parameters. Using only their proper motions and parallaxes, based on a nonlinear rotation model, we found the following parameters of the angular velocity of Galactic rotation:
 $\Omega_0 =28.01\pm0.15$~km s$^{-1}$ kpc$^{-1}$,
 $\Omega^{'}_0=-3.674\pm0.040$~km s$^{-1}$ kpc$^{-2}$ and
 $\Omega^{''}_0=0.565\pm0.023$~km s$^{-1}$ kpc$^{-3}$.
Here, the circular rotation velocity
of the solar neighborhood around the Galactic center is $V_0=226.9\pm3.1$~km s$^{-1}$ for the adopted distance $R_0=8.1\pm0.1$~kpc.

To determine the parameters of the spiral density
wave, we applied a method based on a periodogram
Fourier analysis. This method takes into account
both the logarithmic behavior of the Galactic spiral
pattern and the position angles of objects, which
makes it possible to perform an accurate analysis of
the velocities of objects distributed in a wide range of
Galactocentric distances.

Initially, such an analysis was applied to the sample of 233 youngest OSCs with measured line-of-sight velocities. We showed that the perturbation wavelengths and velocity perturbations found independently for each type of velocities, $\lambda_R=3.3\pm0.5$~kpc and $\lambda_\theta=2.6\pm0.6$~kpc, agree in principle. For the four-armed spiral pattern ($m = 4$ and the adopted $R_0$) the pitch angles $i_R=-14.5\pm2.1^\circ$ and
$i_\theta=-11.4\pm2.6^\circ$ correspond to these values. The Sun’s phase in the spiral density wave is 
 $(\chi_\odot)_R\approx-180^\circ$ and 
 $(\chi_\odot)_\theta\approx-120^\circ$. The amplitudes of the radial
and tangential velocity perturbations are $f_R=9.1\pm0.8$~km s$^{-1}$ and $f_\theta=4.6\pm1.2$~km s$^{-1}$, respectively.

Then, we showed that the influence of the spiral
density wave also manifests itself in the space velocities
of OSCs with ages in the interval 60--300 Myr
(a mean age of 163 Myr). Based on 398 OSCs
from this sample, we performed a spectral analysis
of their radial and residual tangential velocities. The
parameters of the spiral density wave are well determined
from the radial velocities of these OSCs. For
example, the perturbation wavelength and velocity
perturbations were found to be $\lambda_R=2.2\pm0.6$~kpc
($i_R=-9.6\pm2.6^\circ$ for $m = 4$ and the adopted $R_0$),
$f_R=6.1\pm1.8$~km s$^{-1}$ and $f_\theta=3.9\pm2.2$~km s$^{-1}$.
The Sun’s phase in the spiral density wave here is
$(\chi_\odot)_R\approx-90^\circ$ and $(\chi_\odot)_\theta\approx-180^\circ$.

\bigskip \bigskip\medskip{\bf REFERENCES}{\small

1. I. Ablimit, G. Zhao, C. Flynn, and S. A. Bird, Astrophys. J. 895, L12 (2020).

2. L. H. Amaral and J. R. D. L\'epine, Mon. Not. R. Astron. Soc. 286, 885 (1997).

3. C. Babusiaux, F. van Leeuwen, M. A. Barstow, et al. (Gaia Collab.), Astron. Astrophys.
616, 10 (2018).

4. A. T. Bajkova and V. V. Bobylev, Astron. Lett. 38, 549 (2012).

5. V. V. Bobylev, A. T. Bajkova, and A. S. Stepanishchev, Astron. Lett. 34, 515 (2008).

6. V. V. Bobylev and A. T. Bajkova, Mon. Not. R. Astron. Soc. 437, 1549 (2014).

7. V. V. Bobylev and A. T. Bajkova, Mon. Not. R. Astron. Soc. 447, L50 (2015).

8. V. V. Bobylev, A. T. Bajkova, and K. S. Shirokova, Astron. Lett. 42, 721 (2016).

9. V. V. Bobylev and A. T. Bajkova, Astron. Lett. 44, 675 (2018).

10. V. V. Bobylev and A. T. Bajkova, Astron. Lett. 45, 109 (2019).

11. V. V. Bobylev and A. T. Bajkova, Astron. Rep. 65, 498 (2021).

12. V. V. Bobylev, A. T. Bajkova, A. S. Rastorguev, and M. V. Zabolotskikh, Mon. Not. R. Astron. Soc. 502, 4377 (2021).

13. J. Bovy, Mon. Not. R. Astron. Soc. 468, L63 (2017).

14. A. G. A. Brown, A. Vallenari, T. Prusti, et al. (Gaia Collab.), Astron.
Astrophys. 595, 2 (2016).

15. A. G. A. Brown, A. Vallenari, T. Prusti, et al. (Gaia Collab.), Astron. Astrophys.
616, 1 (2018).

16. A. G. A. Brown, A. Vallenari, T. Prusti, et al. (Gaia Collab.), Astron. Astrophys.
649, 1 (2021).

17. W. B. Burton, Astron. Astrophys. 10, 76 (1971).

18. D. Camargo, C. Bonatto, and E. Bica, Mon. Not. R. Astron. Soc. 450, 4150 (2015).

19. T. Cantat-Gaudin, C. Jordi, A. Vallenari, et al., Astron. Astrophys. 618, A93 (2018).

20. T. Cantat-Gaudin, F. Anders, A. Castro-Ginard, et al., Astron. Astrophys.
640, A1 (2020).

21. X.-Q. Cui, Y.-H. Zhao, Y.-Q. Chu, et al., Res. Astron. Astrophys. 12, 1197 (2012).

22. A. K. Dambis, L. N. Berdnikov, Yu. N. Efremov, et al., Astron. Lett. 41, 489 (2015).

23. W. S. Dias, J. R. D. L\'epine, and B. S. Alessi, Astron. Astrophys. 376, 441 (2001).

24. W. S. Dias, M. Assafin, V. Fl\'orio, B. S. Alessi, and
V. Libero, Astron. Astrophys. 446, 949 (2006).

25. W. S. Dias, H. Monteiro, A. Moitinho, et al.,
Mon. Not. R. Astron. Soc. 504, 356 (2021).

26. E. V. Glushkova, A. K. Dambis, A. M. Mel’nik, and A. S. Rastorguev, Astron. Astrophys. 329, 514 (1998).

27. C. J. Hao, Y. Xu, L. G. Hou, et al., Astron. Astrophys. 652, 102 (2021).

28. T. C. Junqueira, C. Chiappini, J. R. D. L\'epine, I. Minchev, and B. X. Santiago, Mon. Not. R. Astron. Soc. 449, 2336 (2015).

29. N. V. Kharchenko, A. E. Piskunov, S. R\"oser,
E. Schilbach, and R.-D. Scholz, Astron. Astrophys. 438, 1163 (2005).

30. N. V. Kharchenko, R.-D. Scholz, A. E. Piskunov,
S. R\"oser, and E. Schilbach, Astron. Nachr. 328, 889 (2007).

31. N. V. Kharchenko, A. E. Piskunov, E. Schilbach,
S. R\"oser, and R.-D. Scholz, Astron. Astrophys. 558, A53 (2013).

32. M. A. Kuhn, L. A. Hillenbrand, A. Sills, E. D. Feigelson,
and K. V. Getman,Astrophys. J. 870, 32 (2018).

33. J. R. D. L\'epine, W. S. Dias, and Yu. Mishurov, Mon.
Not. R. Astron. Soc. 386, 2081 (2008).

34. C. C. Lin and F. H. Shu, Astrophys. J. 140, 646
(1964).

35. L. Lindegren, J. Hernandez, A. Bombrun, et al. (Gaia Collab.), Astron. Astrophys.
616, 2 (2018).

36. A. V. Loktin and G. V. Beshenov, Astron. Rep. 47, 6
(2003).

37. A. V. Loktin and M. E. Popova, Astron. Rep. 51, 364
(2007).
38. A. V. Loktin and M. E. Popova, Astrophys. Bull. 74,
270 (2019).

39. J.Maiz Apell\'aniz, arXiv: 2110.01475 (2021).

40. A. M. Mel’nik, A. K. Dambis, and A. S. Rastorguev,
Astron. Lett. 27, 611 (2001).

41. H. Monteiro, D. A. Barros, W. S. Dias, and J. R. D.
L\'epine, Front. Astron. Space Sci. 8, 62 (2021).

42. P. Mr\'oz, A. Udalski, D. M. Skowron, et al., Astrophys. J. 870, L10 (2019).

43. S. Naoz and N. J. Shaviv, New Astron. 12, 410
(2007).

44. A. E. Piskunov, N. V. Kharchenko, S. R\"oser, E. Schilbach, and R.-D. Scholz, Astron. Astrophys. 445, 545 (2006).

45. M. E. Popova and A. V. Loktin, Astron. Lett. 31, 171 (2005).

46. T. Prusti, J. H. J. de Bruijne, A. G. A. Brown, et al. (Gaia Collab.), Astron.
Astrophys. 595, 1 (2016).

47. A. S. Rastorguev, M. V. Zabolotskikh, A. K. Dambis,
N. D. Utkin, V. V. Bobylev, and A. T. Bajkova, Astrophys.
Bull. 72, 122 (2017).

48. M. J. Reid, K. M. Menten, A. Brunthaler, et al., Astrophys. J. 885, 131 (2019).

49. F. Ren, X. Chen, H. Zhang, R. de Grijs, L. Deng, and Yang Huang, Astrophys. J. Lett. 911, 20 (2021).

50. R.-D. Scholz, N. V. Kharchenko, A. E. Piskunov, S. R\"oser, and E. Schilbach, Astron. Astrophys. 581, A39 (2015).

51. R. Sch\"onrich, J. J. Binney, and W. Dehnen, Mon. Not. R. Astron. Soc. 403, 1829 (2010).

52. Y. Tarricq, C. Soubiran, L. Casamiquela, et al., Astron. Astrophys. 647, A19 (2021).

53. F. Wang, H.-W. Zhang, Y. Huang, et al., Mon. Not. R. Astron. Soc. 504, 199 (2021).

54. M.-S. Xiang, X.-W. Liu, H.-B. Yuan, et al., Mon. Not. R. Astron. Soc. 467, 1890 (2017).

55. M. V. Zabolotskikh, A. S. Rastorguev, and A. K. Dambis, Astron. Lett. 28, 454 (2002).

 }
\end{document}